\documentclass[prl,aps,amsmath,amssymb,showpacs,superscriptaddress,twocolumn]{revtex4-1}

\usepackage{bm,color,float,dcolumn,amssymb,graphicx,hyperref,subfigure}

\begin{document}

\title{Universal entropy of conformal critical theories on a Klein bottle}

\author{Hong-Hao Tu}

\affiliation{Physics Department, Arnold Sommerfeld Center for Theoretical Physics and Center for NanoScience,
Ludwig-Maximilians-Universit\"at M\"unchen, 80333 M\"unchen, Germany}

\affiliation{Institute of Theoretical Physics, Technische Universit\"at Dresden, 01062 Dresden, Germany}

\date{\today}

\begin{abstract}
We show that rational conformal field theories in 1+1 dimensions on a Klein bottle, with length $L$ and width $\beta$, satisfying $L \gg \beta$, have a universal entropy. This universal entropy depends on the quantum dimensions of the primary fields and can be accurately extracted by taking a proper ratio between the Klein bottle and torus partition functions, enabling the characterization of conformal critical theories. The result is checked against exact calculations in quantum spin-1/2 XY and Ising chains.
\end{abstract}

\maketitle

{\em Introduction} -- The characterization of phases and phase transitions is an important task in condensed matter physics. The discovery of topological phases of matter \cite{wen2004}, examples of which include integer and fractional quantum Hall states \cite{klitzing1980,tsui1982}, as well as topological insulators and topological superconductors \cite{hasan2010,qi2011}, greatly enriches our understanding of quantum phases. The topological systems exhibit intriguing behaviors, such as gapless edge states, robust ground-state degeneracy, and quasiparticles with fractional statistics. Because of these remarkable properties, they constitute an important candidate for quantum information processing devices.

Much effort has been made for characterizing these so-called topological phases. In 2+1 dimensions, many topological phases have gapless edge states that are exponentially localized at the boundary. These edge modes are either unidirectional or bidirectional, generically have linear dispersion, and thus are described by chiral or nonchiral conformal field theory (CFT) in 1+1 dimensions. The best-understood examples are fractional quantum Hall states, for which the bulk properties are fully characterized by the edge chiral CFTs through a remarkable bulk-edge correspondence \cite{moore1991}. The nonchiral CFTs can appear as the edge theory of (2+1)-dimensional symmetry-protected topological (SPT) phases, such as the $Z_2$ topological insulators \cite{kane2005} protected by time-reversal and charge conservation symmetries. Needless to say, identifying the edge CFT is an important step toward the full characterization of topological phases in 2+1 dimensions.

In this Letter, we show that (1+1)-dimensional \emph{nonchiral} \emph{rational} CFTs, when placed on a Klein bottle (with length $L$ and width $\beta$, satisfying $L \gg \beta$), have a universal entropy $S=\ln g$. This entropy depends on the quantum dimensions of the CFT primary fields and, therefore, provides a useful quantity which, at least partially, distinguishes different CFTs. This result is directly applicable to (1+1)-dimensional quantum chains and
two-dimensional classical statistical models, when their low-energy effective theories are rational CFTs, and is potentially applicable for (2+1)-dimensional SPT phases with nonchiral gapless edge states. As a first step toward its applications in lattice models, we focus on (1+1)-dimensional quantum chains and devise a \emph{Klein twist} approach to extract the universal entropy. The validity of the Klein twist approach is checked against analytical calculations in (1+1)-dimensional quantum models -- XY and Ising chains. Finally, we discuss the validity and limitations of the Klein twist and briefly comment on a possible way to implement it at the edge of (2+1)-dimensional SPT phases.

\begin{figure}
\centering
\includegraphics[scale=0.27]{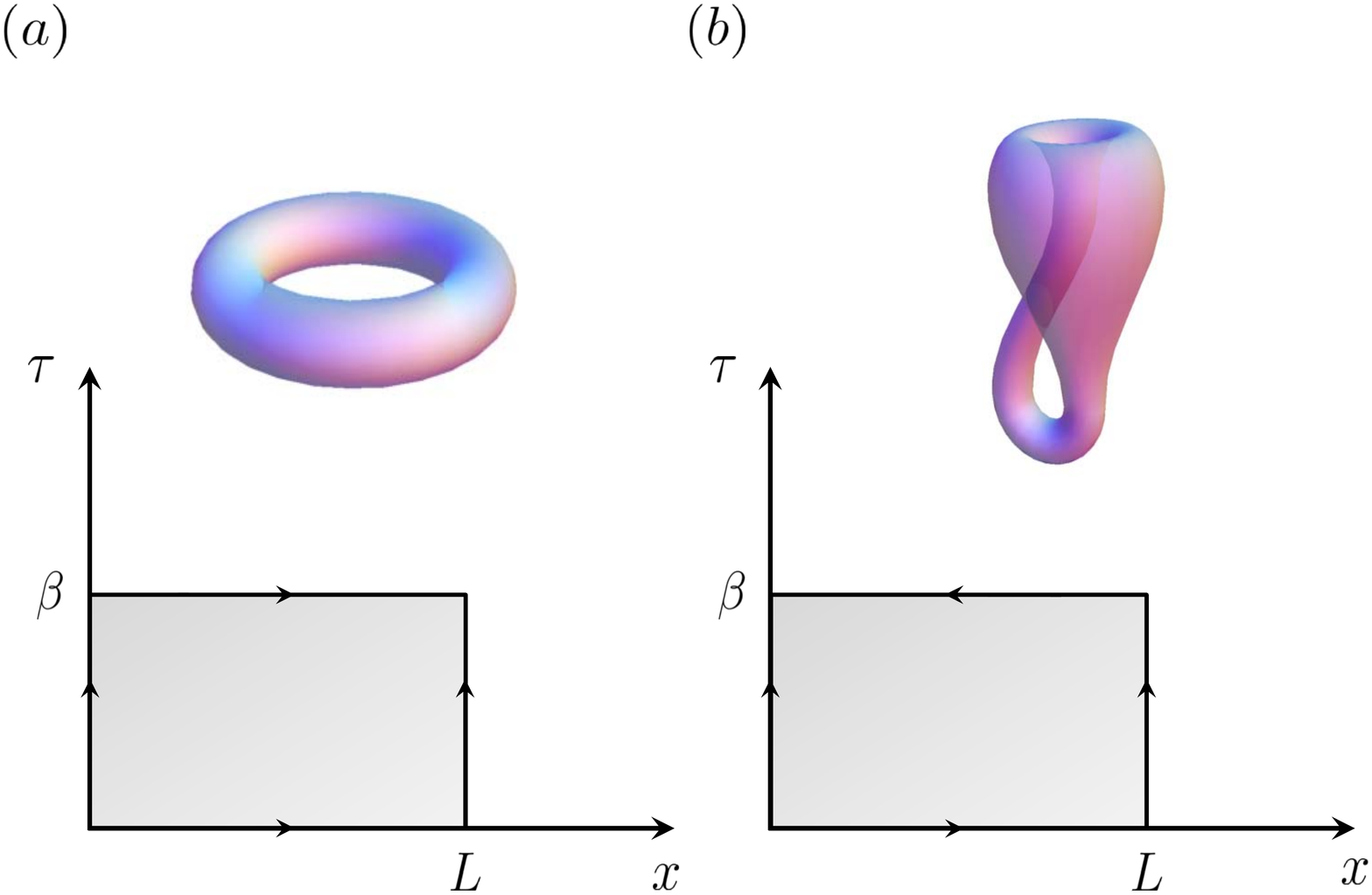}
\caption{(a) Torus and (b) Klein bottle partition functions. The arrows indicate how the opposite edges at the space and imaginary time directions are identified.}
\label{fig:manifold}
\end{figure}

{\em Torus vs. Klein bottle partition functions} --- Let us start with a (1+1)-dimensional quantum system, described by a Hamiltonian $H$, with length $L$ and periodic boundary conditions. At temperature $T=\beta^{-1}$, the partition function $Z^{\mathcal T}=\mathrm{tr}(e^{-\beta H})$, where $\mathcal{T}$ stands for torus, can be cast into an Euclidean path integral on a torus of width $\beta$ in the imaginary time direction and length $L$ in the spatial direction [see Fig.~\ref{fig:manifold}(a)].

When $H$ is critical and is described by a CFT \cite{francesco1997}, the Hamiltonian factorizes at the low-energy and long-wavelength limit into left and right movers with linear dispersion and is written as
\begin{equation}
H = \frac{2\pi v}{L}(L_0 + \bar{L}_0 - \frac{c}{12}) \, ,
\end{equation}%
where $L_0$ ($\bar{L}_0$) is the zeroth-level holomorphic (antiholomorphic) Virasoro generators for the right (left) mover and $c$ and $v$ the central charge and velocity, respectively, of the CFT. The energy eigenstates $|\alpha,\bar{\beta} \rangle$ satisfy $L_0|\alpha, \bar{\beta} \rangle = h_{\alpha}|\alpha, \bar{\beta}\rangle$ and $\bar{L}_0| \alpha, \bar{\beta} \rangle = \bar{h}_{\beta}|\alpha, \bar{\beta} \rangle$, where $(h_{\alpha},\bar{h}_{\beta})$ are conformal weights. Accordingly, the associated Hilbert space is the tensor product of the holomorphic and antiholomorphic sectors, $\mathcal{H} \otimes \bar{\mathcal{H}}$ with $|\alpha\rangle \in \mathcal{H}$ and $|\bar{\beta}\rangle \in \bar{\mathcal{H}}$. The torus partition function is then given by
\begin{equation}
Z^{\mathcal T}
= \mathrm{tr}_{\mathcal{H} \otimes \bar{\mathcal{H}}}(q^{L_0 - \frac{c}{24}} \bar{q}^{\bar{L}_0 - \frac{c}{24}}) \, ,
\label{eq:Ztorus}
\end{equation}%
where $q=e^{2\pi i\tau}$ ($\tau=i v\beta/L$) and $\bar{q}$ is the complex conjugate of $q$.

The Klein bottle partition function takes a similar form (Klein bottle denoted by $\mathcal{K}$) \cite{blumenhagen2009}
\begin{equation}
Z^{\mathcal K}
= \mathrm{tr}_{\mathcal{H} \otimes \bar{\mathcal{H}}}(\Omega q^{L_0 - \frac{c}{24}} \bar{q}^{\bar{L}_0 - \frac{c}{24}}) \, ,
\label{eq:ZKlein}
\end{equation}%
but has an extra operator $\Omega$ inserted. When acting on the tensor product Hilbert space $\mathcal{H} \otimes \bar{\mathcal{H}}$, $\Omega$ interchanges the states in holomorphic and antiholomorphic sectors, $\Omega |\alpha, \bar{\beta}\rangle = |\beta, \bar{\alpha}\rangle$. The physical picture of (\ref{eq:ZKlein}) is that the left and right movers are swapped after the imaginary time evolution and the Euclidean path integral is then glued back (via taking trace) after the swapping. Thus, the Klein bottle partition function is periodic only in the spatial direction but contains a twist in the imaginary time direction  [see Fig.~\ref{fig:manifold}(b)], as opposed to the torus partition function that is periodic in both spatial and time directions. To evaluate (\ref{eq:ZKlein}), a major step is to observe that only left-right symmetric states $|\alpha, \bar{\alpha}\rangle$ have contributions to the trace \cite{blumenhagen2009}, and one is lead to the following partition function that is effectively holomorphic:
\begin{equation}
Z^{\mathcal K}
= \mathrm{tr}_{\mathcal{H_{\mathrm{sym}}}}( q^{ 2L_0 - \frac{c}{12} } ) \, ,
\label{eq:ZKleinSimplified}
\end{equation}%
where $\mathcal{H_{\mathrm{sym}}}$ denotes the subspace of $\mathcal{H} \otimes \bar{\mathcal{H}}$ formed by left-right symmetric states.

The target of the present work is \emph{rational} CFT, a subclass of CFTs whose Hilbert space is organized into a \emph{finite} number of conformal towers. In each tower, the Hilbert space is formed by a primary state and their descendants. A compact way of keeping track of the states in each tower is through the so-called character, defined by $\chi_a(q) = \mathrm{tr}_a(q^{L_0 - \frac{c}{24}})$, where $a$ labels the primary state (and thus also the tower) and the trace is over all states in the tower labeled by $a$. With such a decomposition, the torus partition function in (\ref{eq:Ztorus}) is written as
\begin{equation}
Z^{\mathcal T} = \sum_{a,b} \chi_a(q) M_{a,b} \bar{\chi}_{b}(\bar{q}) \, ,
\label{eq:Ztorus1}
\end{equation}%
where $M_{a,b}$ are non-negative integers specifying the number of distinct primary states $(a,\bar{b})$ in $\mathcal{H} \otimes \bar{\mathcal{H}}$. Accordingly, the Klein bottle partition function in (\ref{eq:ZKleinSimplified}) takes the form of a (weighted) sum of single characters,
\begin{equation}
Z^{\mathcal K} = \sum_{a} M_{a,a} \chi_{a}(q^2) \, .
\label{eq:ZKlein1}
\end{equation}%

In the limit of our interest, $L \gg v\beta$, the partition functions (\ref{eq:Ztorus1}) and (\ref{eq:ZKlein1}) can be evaluated by using the modular transformation properties of the characters, $\chi_{a}(q) = \sum_{b} S_{ab}\chi_b(\tilde{q})$, with $q=e^{-2\pi \frac{v\beta}{L}}$ and $\tilde{q}=e^{-2\pi \frac{L}{v\beta}}$ and $S_{ab}$ the modular $S$ matrix. Since $\tilde{q}\rightarrow 0$, the right-hand side is dominated by the primary state contributions, so that $\chi_{a}(q) \simeq \sum_{b} S_{ab} e^{-2\pi \frac{L}{v\beta}(h_b - \frac{c}{24})}$, where $h_b$ is the conformal weight of the primary state in sector $b$. In the low-temperature limit, $\beta \rightarrow \infty$, the identity sector $I$ with $h_I = 0$ dominates all other sectors with positive $h_b$, which yields $\chi_{a}(q) \simeq S_{aI}e^{2\pi \frac{L}{v\beta} \frac{c}{24}}$. The torus partition function is, after taking into account its modular invariance requirement ($S^{\dag}MS=M$) and the nondegeneracy of the identity sector ($M_{I,I}=1$), given by $Z^{\mathcal T} \simeq e^{\frac{\pi cL}{6v\beta}}$. Similarly, the Klein bottle partition function (\ref{eq:ZKlein1}) can be evaluated by using the modular transformation property of the character, $\chi_{a}(q^2) = \sum_{b} S_{ab}\chi_b(\tilde{q}^{1/2})$, which is justified by replacing $\beta$ by $2\beta$ in the modular transformation used above for the torus case, and we arrive at $Z^{\mathcal K}  \simeq g e^{\frac{\pi cL}{24\beta v}}$, where $g = \sum_a M_{a,a} S_{aI}$. By using the topological quantum field theory terminology, $S_{aI} = d_a/\mathcal{D}$, where $d_a$ is the quantum dimension of the primary field $a$ and $\mathcal{D}$ the total quantum dimension $\mathcal{D}=\sqrt{\sum_a d_a^2}$, we obtain $g=\sum_a M_{a,a} d_a/\mathcal{D}$.

Until now, we have restricted ourselves to pure CFT derivations, from which universal contributions to the partition functions have been obtained. For lattice models, nonuniversal terms generally appear and we expect
\begin{eqnarray}
\ln Z^{\mathcal T}  & \simeq &  -f_0 \beta L + \frac{\pi c}{6v\beta}L  \, ,  \label{eq:Ztorus2} \\
\ln Z^{\mathcal K}  & \simeq &  -f_0 \beta L + \frac{\pi c}{24v\beta}L + \ln g \, , \label{eq:ZKlein2}
\end{eqnarray}%
where the nonuniversal constant, $f_0$, has the meaning of free energy per unit length. When applying to two-dimensional statistical models with spatially isotropic couplings (see, e.g., \cite{lu2001,chui2002}), the velocity is $v=1$. We note that (\ref{eq:Ztorus2}) is the seminal result obtained by Affleck \cite{affleck1986a} and Bl\"ote, Cardy, and Nightingale \cite{blote1986} and (\ref{eq:ZKlein2}) is the key result of the present work.

Several comments are in order regarding (\ref{eq:Ztorus2}) and (\ref{eq:ZKlein2}): (i) $f_0$ takes the same value for torus and Klein bottle cases when $L$ and $\beta$ are large. The reason is that, in the path integral picture, the nonuniversal term $f_0 \beta L$ is determined by the space-time area, while torus and Klein bottle differ only at the ``boundary'' in the time direction. (ii) The second terms are responsible for the specific heat, with $C^{\mathcal K}=\frac{1}{4}C^{\mathcal T} = \frac{\pi c}{12v}T$, which indicates that the heat capacity of a conformal critical system on the Klein bottle is exactly $1/4$ of that for the same system on the torus. (iii) There exists a universal entropy $\ln g$ for conformal critical systems on the Klein bottle. The origin of this entropy, from a technical point of view, is analogous to the celebrated Affleck-Ludwig entropy \cite{affleck1991a} for boundary CFTs: They share the same feature that the partition functions are given by a sum of single characters, from which the universal entropies arise when performing the modular transformation. In this sense, the universal entropy on the Klein bottle, which by itself is a closed manifold, may be viewed as a ``boundary'' entropy without boundary. Very recently, Ref.~\onlinecite{tang2017} transformed the Klein bottle into a cylinder with nonlocal interactions along the time direction and attributed the universal entropy to the nonlocal boundary interactions on the cylinder, which thus provides an appealing physical picture for the origin of the Klein bottle universal entropy and its connection to the Affleck-Ludwig entropy. (iv) The following ratio between the Klein bottle and torus partition functions is universal:
\begin{equation}
\frac{Z^{\mathcal K}(2L,\frac{\beta}{2})}{Z^{\mathcal T}(L,\beta)} = g \, .
\label{eq:Zratio}
\end{equation}
As $g$ depends on the operator content of the CFT (via $g=\sum_a M_{a,a} d_a/\mathcal{D}$), it is, in general, a noninteger and quantifies the universal, intensive ``ground-state degeneracy'' of rational CFTs on the Klein bottle. Since the value of $g$ is related to the quantum dimensions of the primary fields, it provides a useful label which, at least partially, distinguishes different CFTs. (v) The partition function ratio in (\ref{eq:Zratio}) may also be used for detecting and locating phase transitions occurring at zero temperature: (1+1)-dimensional quantum critical models are mostly described by CFTs with diagonal partition function, i.e., $M_{a,b}=\delta_{ab}$ in (\ref{eq:Ztorus1}), and thus $g=\sum_a d_a/\mathcal{D} \geq 1$ ($g=1$ is achieved only for very special CFTs with a single primary, such as $E_8$ level 1 CFT). When entering a gapped phase, say, with a unique ground state separated from excitations by a gap $\Delta$, the area-proportional terms would dominate in (\ref{eq:Ztorus2}) and (\ref{eq:ZKlein2}), while all other terms are exponentially suppressed by the gap at a low temperature ($T \ll \Delta$), and then one obtains $g=1$. Thus, tuning the coupling constant in a Hamiltonian from a gapped phase to a critical point (described by the CFT) would indicate a sharp change of $g$ (see Ref.~\cite{chen2017} for a detailed analysis of an explicit example).


{\em XY and Ising chains} --- Now we focus on two quantum spin-1/2 chains, i.e., the XY and Ising models, for which exact calculations of the torus and Klein bottle partition functions can be performed. More importantly, they shed light on how Klein bottle partition functions may be constructed and justified for generic lattice systems.

Both models consider spin-1/2 particles on periodic chains with $L$ sites ($L$ even), the Hamiltonians of which are given by
\begin{equation}
H_{\mathrm{XY}} = -\sum_{j=1}^{L}(\sigma^x_j \sigma^x_{j+1} + \sigma^y_j \sigma^y_{j+1}) \, ,
\label{eq:XYHamiltonian}
\end{equation}
and
\begin{equation}
H_{\mathrm{Ising}} = -\sum_{j=1}^{L}(\sigma^x_j \sigma^x_{j+1} + \sigma^z_{j}) \, ,
\label{eq:IsingHamiltonian}
\end{equation}
where $\sigma^\nu$ ($\nu=x,y,z$) are Pauli matrices and $\sigma^\nu_{L+1} = \sigma^\nu_1$.

The torus partition functions are, of course, defined in a usual way: $Z^{\mathcal T} = \mathrm{tr}(e^{-\beta H})$. For the Klein bottle partition function, the nontrivial task is to find an operator defined on the lattice, which, when acting on the low-energy states, plays the role of interchanging left and right movers, as required in the definition of $\Omega$ in (\ref{eq:ZKlein}). For the XY and Ising chains with even $L$, we have proven that it is simply the bond-centered reflection operator $P$, defined by
\begin{equation}
P |s_1, s_2, \ldots, s_{L-1}, s_L \rangle = |s_L, s_{L-1}, \ldots, s_2, s_1 \rangle \, ,
\label{eq:reflection}
\end{equation}
which plays that role. Here $|s_j\rangle$ denotes the spin state at site $j$, $s_j=\pm 1$. The Klein bottle partition functions are hence given by $Z^{\mathcal K} = \mathrm{tr}(Pe^{-\beta H})$. We name such a construction of $Z^{\mathcal K}$ by inserting a lattice reflection as the \emph{Klein twist}, whose connection to the Klein bottle becomes transparent when representing $e^{-\beta H}$ in $Z^{\mathcal K}$ by using the Trotter-Suzuki decomposition.

The justification of the Klein twist is most conveniently done in the energy eigenstate basis, for which we need to diagonalize the Hamiltonian. Both XY and Ising chains can be solved via the Jordan-Wigner transformation, $\sigma^x_j=(c^{\dag}_j+c_j)(-1)^{\sum_{l<j}n_l}$ and $\sigma^z_j=2n_j-1$ with $n_j=c^{\dag}_j c_j$, which map spins to fermions. Below, we illustrate such a calculation for the XY chain and the analysis for the Ising chain requires only minor modifications. When taking into account that the fermion parity $Q = (-1)^{\sum_{j=1}^L n_j}$ is a conserved quantity, the Hilbert space splits into two sectors with definite fermion parity $Q=\pm 1$, and one arrives at \cite{lieb1961,katsura2011}
\begin{equation}
H_{\mathrm{XY}} = \frac{1+Q}{2} H_{\mathrm{XY}}^+ + \frac{1-Q}{2} H_{\mathrm{XY}}^- \, ,
\label{eq:fermionicXY}
\end{equation}
where $H_{\mathrm{XY}}^{\pm}=-2\sum_{j=1}^{L}(c^{\dag}_j c_{j+1} + c^{\dag}_{j+1} c_j)$ with $c_{L+1}= \mp c_1$. The two sectors, following the CFT convention, are termed as the Neveu-Schwarz and Ramond sectors, respectively. The Hamiltonian in the Neveu-Schwarz sector is then diagonalized in momentum space, $H_{\mathrm{XY}}^+= \sum_p \varepsilon_p c^{\dag}_p c_p $, where $\varepsilon_p = -4\cos p$, $c_p=\frac{1}{\sqrt{L}}\sum_j c_j e^{-ipj}$, and the allowed lattice momenta are $p=\pm\frac{\pi}{L},\pm\frac{3\pi}{L},\ldots,\pm\frac{\pi(L-1)}{L}$. The Hamiltonian in the Ramond sector takes the same form, while the allowed lattice momenta are instead $q=0,\pm\frac{2\pi}{L},\ldots,\pm\frac{\pi(L-2)}{L},\pi$. The complete energy eigenstate basis is formed by creating an even (odd) number of fermionic modes on top of the vacuum $|0\rangle$ in the Neveu-Schwarz (Ramond) sector. When working in this basis, the torus partition function can be immediately written down as
\begin{eqnarray}
Z^{\mathcal T}_{\mathrm{XY}} &=& \frac{1}{2} \prod_p (1+e^{-\beta \varepsilon_p}) + \frac{1}{2} \prod_p (1-e^{-\beta \varepsilon_p}) \nonumber \\
&\phantom{=}& + \frac{1}{2} \prod_q (1+e^{-\beta \varepsilon_q})\, .
\label{eq:ZXYtorus}
\end{eqnarray}

The critical theory for describing the XY chain is known to be the $U(1)_4$ CFT of a free boson with central charge $c=1$. This CFT has four primary fields, $I, \, s, \, \bar{s}, \, v$ with conformal dimensions $h_{I}=0$, $h_{s}=h_{\bar s}=1/8$ and $h_{v}=1/2$. The correspondence of (\ref{eq:ZXYtorus}) to the CFT torus partition function (\ref{eq:Ztorus1}) can be made clear by linearizing the spectrum of fermions close to two Fermi points, $k_{\mathrm F} = \pm\pi/2$. It is straightforward to see that the resulting CFT partition function is diagonal, $Z^{\mathcal T}_{\mathrm{XY}} \simeq \sum_{a=I,s,\bar{s},v} \chi_a(q) \bar{\chi}_{a}(\bar{q})$, and the Neveu-Schwarz (Ramond) sector corresponds to primary fields $I, \, v$ ($s, \, \bar{s}$), and their descendants, respectively.

We are now in a position to justify the usage of (\ref{eq:reflection}) for constructing the lattice version of the Klein bottle partition function for the XY chain (\ref{eq:XYHamiltonian}). Our strategy is to show that the contributions to $Z^{\mathcal K}$ come from left-right symmetric states (in the lattice sense), which agrees with the CFT derivation. To achieve this, we need to work out the action of the reflection operator $P$ on the energy eigenstate basis. We note that the action of the reflection operator on spins, $P \sigma^{\nu}_j P^{-1} = \sigma^{\nu}_{L-j+1}$, is inherited by its action on fermions, $P c^{\dag}_j P^{-1} = c^{\dag}_{L-j+1}Q$, which, in momentum space, is written as
\begin{equation}
P c^{\dag}_p P^{-1} = -e^{ip }c^{\dag}_{-p} Q \, , \quad
P c^{\dag}_q P^{-1} = e^{iq }c^{\dag}_{-q} Q \, .
\label{eq:mode-reflection}
\end{equation}
By using (\ref{eq:mode-reflection}) and noticing that the fermionic vacuum $|0\rangle$ (fully polarized spin-down state) is invariant under reflection, $P|0\rangle=|0\rangle$, one realizes that only a few energy eigenstates survive in the trace $\mathrm{tr}(Pe^{-\beta H})$, since most of the states are orthogonal to their reflected partners. In the Neveu-Schwarz (Ramond) sector, all contributing states can be obtained by starting from the vacuum $|0\rangle$ (singly occupied states $c^{\dag}_{q=0}|0\rangle$ and $c^{\dag}_{q=\pi}|0\rangle$) and creating pairs of fermions with opposite momenta, such as $c^\dag_{-p} c^\dag_p |0\rangle$ ($c^\dag_{-q} c^\dag_qc^{\dag}_{q=0}|0\rangle$ and $c^\dag_{-q} c^\dag_qc^{\dag}_{q=\pi}|0\rangle$). We note that those states originated from $|0\rangle$ and $c^{\dag}_{q=0}|0\rangle$ are invariant under the reflection, while those originated from $c^{\dag}_{q=\pi}|0\rangle$ are invariant up to a minus sign. When taking these into account, the analytical form of the Klein bottle partition function for the XY chain can be obtained as follows:
\begin{eqnarray}
Z^{\mathcal K}_{\mathrm{XY}} &=& \prod_{0<p<\pi} (1+e^{-2\beta \varepsilon_p}) + (e^{-\beta \varepsilon_{q=0}} - e^{-\beta \varepsilon_{q=\pi}}) \nonumber \\
&\phantom{=}& \times  \prod_{0<q<\pi} (1+e^{-2\beta \varepsilon_q}) \, .
\label{eq:ZXYKlein}
\end{eqnarray}

\begin{figure}
\centering
\includegraphics[scale=0.55]{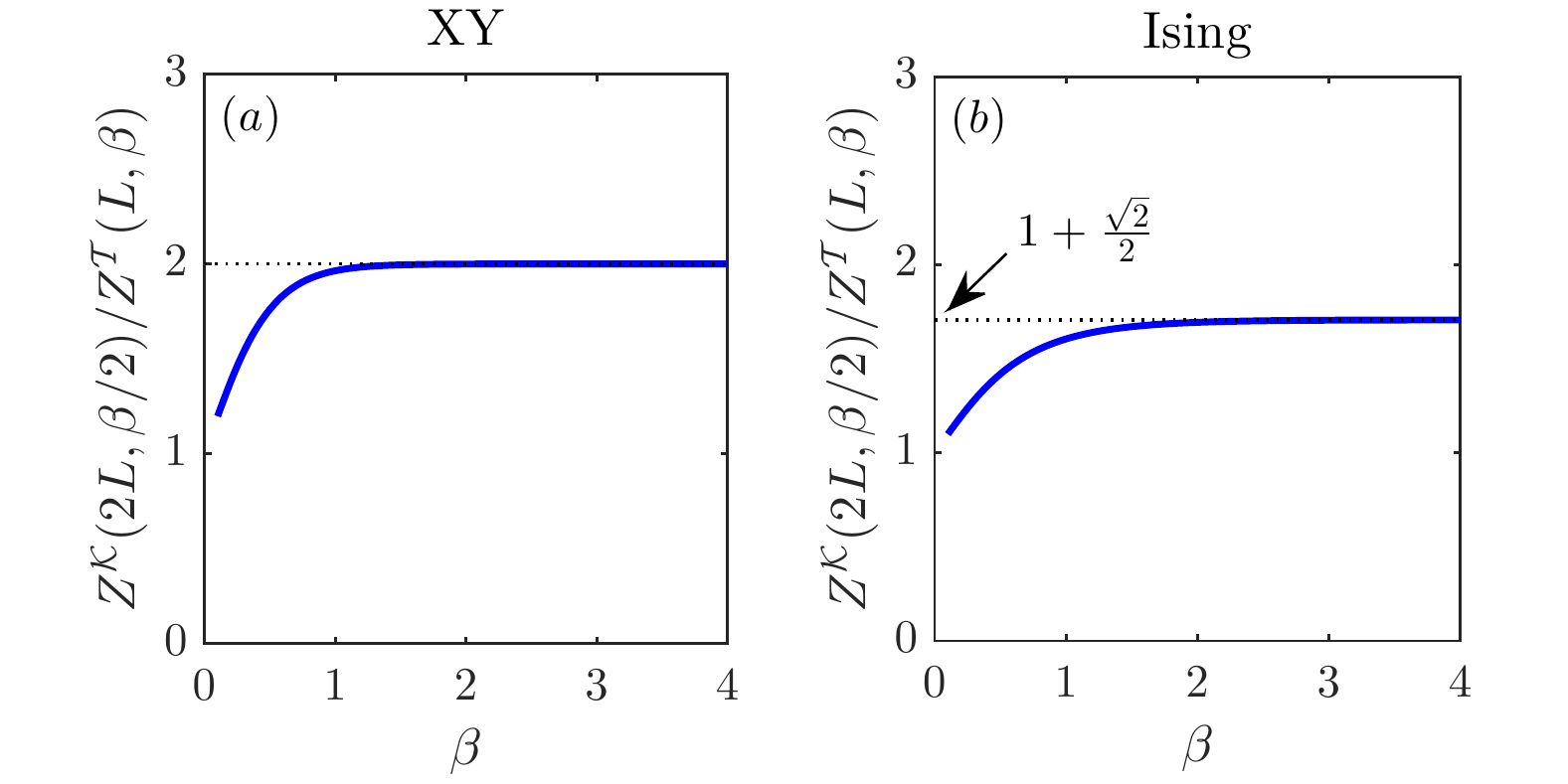}
\caption{Ratio of Klein bottle and torus partition functions $Z^{\mathcal K}(2L,\beta/2)/Z^{\mathcal K}(L,\beta)$ as a function of the inverse temperature $\beta$ for (a) XY and (b) Ising chains with $L=100$. The ratio rapidly converges at low temperature to the CFT predicted values $g_{\mathrm{XY}}=2$ and $g_{\mathrm{Ising}}=1+\frac{\sqrt 2}{2}$.}
\label{fig:kleinentropy}
\end{figure}

When linearizing the fermion spectrum in (\ref{eq:ZXYKlein}), only a single Fermi point $k_{\mathrm F} = \pi/2$ needs to be considered, so that the resulting form is the sum of single characters, $Z^{\mathcal K}_{\mathrm{XY}} \simeq \sum_{a=I,s,\bar{s},v} \chi_a (q^2)$, which is indeed consistent with the CFT result (\ref{eq:ZKlein1}). In Fig.~\ref{fig:kleinentropy}(a), we plot the exact results of the ratio (\ref{eq:Zratio}) for the XY chain with $L=100$. This is in perfect agreement with the CFT prediction: The four primary fields of the $U(1)_4$ CFT are Abelian and thus have quantum dimensions $d_I=d_s=d_{\bar s}=d_v=1$ (total quantum dimension $\mathcal D = 2$), leading to $g_{\mathrm{XY}}=2$.

For the Ising chain, a similar calculation has been performed, and the results are shown in Fig.~\ref{fig:kleinentropy}(b). The extracted value of $g$ from the lattice calculation again agrees very well with the CFT prediction: For the Ising CFT, there are three primary fields $I, \, \sigma, \, \, \psi$ with quantum dimensions $d_{I}=d_{\psi}=1$ and $h_{\sigma}=\sqrt{2}$ (total quantum dimension $\mathcal D = 2$), yielding $g_{\mathrm{Ising}}=1+\sqrt{2}/2$.

The Klein twist approach has also been successfully verified in other quantum chains, such as the $Z_3$ Potts \cite{tang2017} and spin-1 Blume-Capel models \cite{chen2017}. However, the validity of the Klein twist approach deserves special attention. From the above XY and Ising examples, one may notice that the lattice reflection plays two roles: (i) selecting certain states at (many-body) lattice momenta $0$ and $\pi$ and removing all other ``unwanted'' states; (ii) the kept states relevant for low-energy physics are eigenstates of the reflection operator with the \emph{same} eigenvalue (i.e., they share the same parity quantum number $\pm 1$). For models whose gapless modes appear at other lattice momenta, one may need to use combinations of lattice reflection and translation operator, in order to project onto the relevant states. However, a more subtle question is, when the low-energy states have different parity quantum numbers, how to construct the Klein bottle partition function on a lattice. So far, we do not have a satisfactory answer to this question, and it will be left for further investigations.

{\em Conclusion and discussion} --- In summary, we have shown that (1+1)-dimensional nonchiral rational CFTs exhibit a universal entropy on a Klein bottle. This entropy depends on the quantum dimensions of the primary fields and characterizes the CFTs. For (1+1)-dimensional quantum lattice models, we have devised a Klein twist procedure to extract such universal entropy, which found excellent agreement with CFT predictions for quantum XY and Ising chains.

There is no doubt that (2+1)-dimensional SPTs with gapped bulk and nonchiral gapless edges provide an interesting class of system to investigate whether the Klein twist might be useful for identifying the edge CFTs. Recently, it has been shown \cite{cirac2011} that, for (2+1)-dimensional SPT wave functions, a tensor network formulation allows us to represent the thermal density operator $e^{-\beta H_{\mathrm{edge}}}$ of its (1+1)-dimensional gapless edge theory as a matrix-product operator (MPO). Once this MPO is obtained, the Klein twist approach developed here could be used directly.

For future investigations, one interesting question is to understand possible connections of the Klein twist developed in the present work and similar partial twists \cite{shapourian2017} in topological wave functions. Another intriguing issue is to study whether there exists a relation between the Klein bottle universal entropy and the \emph{bulk} renormalization group flow (the Affleck-Ludwig boundary entropy decreases during the \emph{boundary} renormalization group flow \cite{affleck1991a,friedan2004}, but cannot be used for indicating \emph{bulk} renormalization group flows \cite{green2008}).

{\em Acknowledgment} --- We are grateful to Meng Cheng, Xiao-Liang Qi, Germ\'an Sierra and, in particular, Roberto Bondesan, J\'er\^ome Dubail,
Wei Li, Wei Tang, and Lei Wang for stimulating discussions. The support from the DFG through the Excellence Cluster ``Nanosystems Initiative Munich" is acknowledged.

\bibliography{Klein}

\end{document}